\documentclass[10pt,twocolumn,twoside]{IEEEtran}
\usepackage[all]{xy}
\usepackage{color}
\usepackage{algorithm}
\usepackage{algpseudocode}
\usepackage{subfigure}
\usepackage{bm}
\usepackage{graphicx}
\usepackage{color}
\usepackage{wrapfig}
\usepackage{epsf}
\usepackage{times,mathptm}
\usepackage{url}
\usepackage{amssymb, amsmath, mathtools, amsthm}
\usepackage{nomencl}
\usepackage{comment}
\let\labelindent\relax
\usepackage{enumitem}
\makenomenclature
\newtheorem{theorem}{Theorem}

\newtheorem{lemma}{Lemma}
\definecolor{RED}{rgb}{0.6,0.,0.}
\definecolor{BLUE}{rgb}{0.,0.,0.6}
\definecolor{GREEN}{rgb}{0.,0.6,0.}
\definecolor{MALINA}{rgb}{0.6,0.,0.6}
\definecolor{YELLOW}{rgb}{0.8,0.8,0}

\newcommand{\indep}{\rotatebox[origin=c]{90}{$\models$}}

\begin{document}
\title{Topology Estimation using Graphical Models in Multi-Phase Power Distribution Grids}
\author{\IEEEauthorblockN{Deepjyoti~Deka*, Michael~Chertkov*\dag, and Scott~Backhaus*\\}
\IEEEauthorblockA{* Los Alamos National Laboratory, New Mexico, USA\\ \dag Skolkovo Institute of Science and Technology, Moscow, Russia}
\thanks{Deepjyoti Deka and Scott Backhaus are with Los Alamos National Laboratory, Los Alamos, NM 87544. Email: deepjyoti@lanl.gov (corresponding author), backhaus@lanl.gov}
\thanks{M. Chertkov is with Los Alamos National Laboratory, Los Alamos, NM and Skolkovo Institute of Science and Technology, Russia. Email: chertkov@lanl.gov}
\thanks{This work was supported by U.S. Department of Energy’s Grid Modernization Initiative and the Center for Non Linear Studies at Los Alamos.}}
\maketitle
\begin{abstract}
Distribution grid is the medium and low voltage part of a large power system. Structurally, the majority of distribution networks operate radially, such that energized lines form a collection of trees, i.e. forest, with a substation being at the root of any tree. The operational topology/forest may change from time to time, however tracking these changes, even though important for the distribution grid operation and control, is hindered by  limited real-time monitoring. This paper develops a learning framework to reconstruct radial operational structure of the distribution grid from synchronized voltage measurements in the grid subject to the exogenous fluctuations in nodal power consumption. To detect operational lines our learning algorithm uses conditional independence tests for continuous random variables that is applicable to a wide class of probability distributions of the nodal consumption and Gaussian injections in particular. Moreover, our algorithm applies to the practical case of  unbalanced three-phase power flow. Algorithm performance is validated on AC power flow simulations over IEEE distribution grid test cases.
\end{abstract}
\begin{IEEEkeywords}
Distribution networks, Power flow, Unbalanced three-phase, Graphical models, Conditional independence, Computational complexity
\end{IEEEkeywords}

\section{Introduction}
\label{sec:intro}
The operation of a large power grid is separated into different tiers/levels: transmission grid that consists of high voltage lines connecting the generators to the distribution substations, and distribution grid consisting of the medium and low voltage lines that connect distribution substations to loads. The structure of the transmission grid is made loopy to reliable delivery of electricity to substations via multiple, redundant paths. On the other hand, typical distribution grids are operationally radial (tree-like) with the substation at the root and loads positioned along the non-root nodes of the tree. The radial topology is selected from a subset of power distribution lines, which overall i.e. structurally form a loopy graph, by switching on/off breakers \cite{dekatcns}. An illustration of the radial operational topology is presented in Fig.~\ref{fig:city}. Topology estimation in the distribution grid thus refers to the problem of determining the current set of operational lines which are energized, i.e. with respective switch statuses `on'. These changes may conducted in an ad-hoc way without proper and timely reporting to the distribution system operator. Accurate topology estimation in the distribution grid is necessary for checking system status/integrity, e.g. failure detection, and also for taking consecutive optimization and control decisions. Real-time topology estimation is hindered by the limited presence of real-time line-based measurements (flow and/or breaker statuses) in the distribution grid \cite{hoffman2006practical}. 
There has been an emerging effort in providing better observability at the distribution level through deployment of advanced nodal measurement devices like Phasor Measurement Units (PMUs) \cite{phadke1993synchronized}, micro-PMUs \cite{micropmu}, Frequency Monitoring Network (FNET) \cite{fnet} systems and alike. 
Such devices and smart meters  are capable of providing real time measurements of voltages and frequency, though often limited to the grid nodes. The goal of this paper is to utilize nodal measurements of voltages, captured e.g. by smart meters, to efficiently estimate the operational radial topology of the distribution grids.


\subsection{Prior Work}
Topology learning, in power grids in general and in distribution grids in particular, is a fast growing area of research. Past research efforts in this area differ by the methodology for edge detection, available measurements and the types of the flow model used. 
For linearized power flow models in radial grids with constant $R/X$ (resistance to reactance) ratio, \cite{bolognani2013identification} presents a topology identification algorithm using the sign of inverse covariance matrix. \cite{dekatcns,dekaecc,dekasmartgridcomm} present greedy structure learning algorithm using trends in second order moments of nodal voltage magnitudes for linearized power flow models, where observations are known only at a subset of the grid nodes. Sets of line flow measurements have been used for topology estimation using maximum likelihood tests in \cite{ramstanford}. Further, there have been data-driven efforts to identify topology and phase in distribution grids. \cite{berkeley,reno,arya} include machine learning schemes that compare available time-series observations from smart meters with database of permissible signatures to identify topology changes, phase recovery and parameter estimation. 
A group lasso and graph lasso based approximate scheme for topology identification in loopy grids is discussed in \cite{ram_loop, dekairep}. In contrast to prior work, a majority of distribution grids are unbalanced \cite{kerstingbook} and have three-phase voltages and injections at different nodes. The overarching goal of this work is to present an efficient algorithm for topology estimation using multi-phase nodal voltages from a radial distribution grid using the graphical model framework.

\subsection{Contribution}
In this paper, we consider multi-phase power flows in a distribution grid and aim to estimate the radial operational topology using measurements of nodal voltages. The first contribution of this work is the development of a linearized coupled three-phase power flow model that generalizes our prior work in single-phase power flow model \cite{dekatcns} and also relates to similar linearized models specific to radial grids \cite{lowlinear,chen1991distribution}. 
The second contribution of this work consists of proving that under standard assumptions about fluctuations of power consumption the distribution of nodal voltages can be described by a specific chordal graphical model. We show that the edges in the graphical model include both the actual operational lines/edges of the grid-graph  as well as additional edges relating two-hop neighbors in the grid-graph. Based on this factorization of the voltage distribution, we present our learning algorithm that uses conditional independence based tests ($4$ nodes per test, which we also call ``quartet") to identify the operational edges.

Our learning framework has several computational and practical advantages. First, the framework is independent of the exact probability distribution for each individual node's power usage and voltage profile, and hence applicable to general distributions. Second, it does not require knowledge of line impedances and similar network parameters that are calibrated infrequently and hence may not be known accurately. Third, the computational complexity of each edge detection test does not grow with the network size - that is the test is local. Each test considers only a set of four nodes. Furthermore, the tests can be conducted in a distributed fashion. These results extend what was reported earlier in a conference version \cite{dekapscc} where a single-phase version of this paper's results were presented with limited simulations. To the best of our knowledge, this is the first work claiming a provable topology learning algorithm for three-phase distribution grids.

The material in the manuscript is organized as follows. The next, technical introduction,  section provides a brief discussion of the distribution grid topology, power flow models and associated nomenclature. The linearized three-phase power flow model and its single-phase counterpart are described in Section \ref{sec:linearized}. Section \ref{sec:graphicalmodel} analyzes the graphical model of power grid voltage measurements (we emphasize that the induced graphical model is built from but different from the grid-graph itself). Conditional independence properties of the voltage distribution are introduced and utilized in Section \ref{sec:learning} to develop our main learning algorithm. Experimental results on the algorithm's IEEE radial networks test are presented in Section \ref{sec:conclusion}. The last section is reserved for conclusions.

\section{Distribution Grid: Structure \& Power Flows}
\label{sec:structure}
\subsection{Structure}
\begin{figure}[!bt]
\centering
\includegraphics[width=0.30\textwidth]{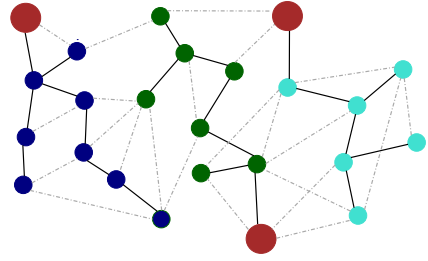}
\vspace{-.25cm}
\caption{Schematic of a radial distribution grid with substations represented by large red nodes. The operational grid is formed by solid lines (black). Load nodes within each tree are marked with the same color.
\label{fig:city}}
\end{figure}

\textbf{Radial Structure}: The distribution grid is represented by a radial graph ${\cal G}=({\cal V},{\cal E})$, where ${\cal V}$ is the set of buses/nodes of the graph and ${\cal E}$ is the set of undirected lines/edges. The operational edge set $\cal E$ is realized by closing switches in an over-complete set of functional lines/edges ${\cal E}_{full}$ (see Fig.~(\ref{fig:city})). The operational grid (note a difference with the functional grid-graph) is a collection of $K$ disjoint trees, $\cup_{k=1,\cdots,K}{\cal T}_k$, where each tree ${\cal T}_k$ spans a subset of the nodes ${\cal V}_{{\cal T}_k}$ with a substation at the root node and connected by the set of operational edges ${\cal E}_{{\cal T}_k}$. We denote nodes by letters  $i$, $j$ and so on. The undirected ledge (line) connecting nodes $i$ and $j$ is denoted by $(ij)$. The non-substation terminal nodes with degree one are called `leaf' nodes. Node connected to a leaf node is called its `parent' node. Nodes with degree greater than $1$ are called `non-leaf' nodes. Path ${\cal P}_{ij}:i-\pi_1-\pi_2-..\pi_n-j$ denotes the unique set of nodes such that edges $(i\pi_1),(\pi_1\pi_2), ..(\pi_nj)$ connect $i$ and $j$.
To circumvent notational complications we limit in the remainder of the manuscript our topology learning problem to grids containing only one operational tree ${\cal T} = {\cal G} = ({\cal V},{\cal E})$. (Extension to the case of an operational forest layout is straightforward.) However, prior to discussing the learning, we will first review in the next subsection  AC power flow models on a general graph-grid (which is not necessarily a tree).

\subsection{Power Flow models}
\textbf{Notations:} Real-/complex- valued quantities are represented in lower/upper case. We use hat- notation for a vector variable with components in multiple phases. The per phase component is described without the hat, with a particular phase as a superscript. The value of a variable at a specific grid location (bus or line) is marked by a subscript. If no subscript is mentioned, it refers to the vector of values for the variable at all permissible locations. For example, $\hat{W}_i$ ($\hat{w}_i$) represents a multi-phase complex (real) variable at location $i$, while $W^a_i$ ($w^a_i$) represents its value at the node $i$ and phase $a$. $W^a$ ($w^a$) represents the vector with complex (real) values at all locations, for phase $a$. 

\textbf{Single-Phase Power Flow (AC-PF)}: Here the voltages, currents and injections in the entire grid are defined over a single-phase (say phase $a$) that is skipped from notation in this paragraph notations for convenience. According to the  Kirchhoff's laws, the complex valued AC-PF equations governing power flow leaving node $i$ of the grid-graph ${\cal G} = ({\cal V},{\cal E})$ is given by:
\begin{align}
\forall i\in{\cal V}:\quad P_i=p_i+\tilde{i} q_i=\sum_{j:(ij)\in  {\cal E}}\frac{v_i^2-v_i v_j e^{\tilde{i}\theta_i-\tilde{i}\theta_j}}{Z_{ij}^*}
\label{p-complex}
\end{align}
where the real valued scalars, $v_i$, $\theta_i$, $p_i$ and $q_i$ denote, respectively, the voltage magnitude, phase, active and reactive power injection at node $i$. $V_i (= v_i\exp(\tilde{i}\theta_i))$ and $P_i$ mark, respectively, the nodal complex voltage and injection ($\tilde{i}^2=-1$). $Z_{ij}=r_{ij}+\tilde{i} x_{ij}$  is the impedance of the  line $(ij)$. $r_{ij}$ ($x_{ij}$) stands for the line resistance (reactance). The unbalanced power flow equations over three phases are described next.

\textbf{Three-Phase Unbalanced Power Flow (AC-PF3) \footnote{Note that even though we discuss solely a three-phase system, our analysis extends to a general $m$ phase system. }}: In a real-world setting, AC power systems have three unbalanced phases that operate over four wires (one for each phase and a common ground). Consequently, the active and reactive power injections at each bus are not scalars but have $3$ components each. Similarly, the complex voltage (magnitude and phase) at each bus and the power flows on each line are of sizes $3 \times 1$, with a component for each phase. Using the notation described earlier, the three-phase complex power injections ($\hat{P}$) and complex voltages ($\hat{V}$) in the grid are
\begin{align}
\forall i\in{\cal V}:\hat{P}_i=\begin{bmatrix}P^{a}_i\\ P^{b}_i\\ P^{c}_i\end{bmatrix} =\begin{bmatrix}p^{a}_i\\p^{b}_i\\p^{c}_i\end{bmatrix}+\tilde{i} \begin{bmatrix}q^{a}_i\\q^{b}_i\\q^{c}_i\end{bmatrix}, \hat{V}_i= \begin{bmatrix}V^{a}_i\\V^{b}_i\\V^{c}_i\end{bmatrix}=\begin{bmatrix}v^{a}_ie^{\tilde{i}\theta^{a}_i}\\v^{b}_ie^{\tilde{i}\theta^{b}_i}\\v^{b}_ie^{\tilde{i}\theta^{c}_i}\end{bmatrix} \nonumber
\end{align}
Here, $[\theta^{a}_i~\theta^{b}_i~\theta^{c}_i]^T$ stands for the voltage phase angles at node $i$. Reference phase angles in the three phase system (all defined modulo $2\pi$) are
\begin{align}
    \hat{\theta}_{ref}=[0~2\pi/3~-2\pi/3]^T \label{theta_ref}
\end{align}
Impedance on line $(ij)$ is represented by a $3\times 3$ symmetric matrix $\hat{Z}$ which relates the three phase current $\hat{I}_{ij}$ over line $(ij)$ to the voltage difference according to
\begin{align}
\hat{Z}_{ij}\hat{I}_{ij} &= \hat{Z}_{ij}[I^{a}_{ij}~ I^{b}_{ij}~ I^{c}_{ij}]^T =  (\hat{V}_{i} - \hat{V}_j) \label{Pline-threephase}\\
\text{where~}{\huge \hat{Z}_{ij}} &= {\huge \hat{r}_{ij}} + \tilde{i}{\huge \hat{x}_{ij}} = \begin{bmatrix} r^{a a}_{ij}~~ r^{a b}_{ij}~~  r^{a c}_{ij}\\r^{b a}_{ij}~~ r^{b b}_{ij}~~  r^{b c}_{ij}\\r^{c a}_{ij}~~ r^{c b}_{ij}~~  r^{c c}_{ij}\end{bmatrix}
+ \tilde{i}\begin{bmatrix} x^{a a}_{ij}~~ x^{a b}_{ij}~~  x^{a c}_{ij}\\x^{b a}_{ij}~~ x^{b b}_{ij}~~  x^{b c}_{ij}\\x^{c a}_{ij}~~ x^{c b}_{ij}~~  x^{c c}_{ij}\end{bmatrix} \nonumber
\end{align}
Note that the diagonal values in $\hat{Z}_{ij}$ denote the in-phase impedances, while the off-diagonal values denote the inter-phase impedances of the line. The real-valued resistance ($\hat{r}_{ij}$) and reactances ($\hat{x}_{ij}$) have a similar structure. The Kirchoff laws in three phases (AC-PF3) for the configuration $\cal F$ are then given by:
\begin{align}
\forall i\in{\cal V}: \hat{P}_i=\smashoperator[lr]{\sum_{j:(ij)\in{\cal E}}}diag\left(\hat{V}_i\hat{I}_{ij}^H\right) =\smashoperator[lr]{ \sum_{j:(ij)\in{\cal E}}}diag\left(\hat{V}_i(\hat{V}^H_{i} - \hat{V}^H_j){\hat{Y}_{ij}}\right)\label{p-threephasecomplex}
\end{align}
where $\hat{Y}_{ij}$, the inverse of $\hat{Z}_{ij}^{H}$, is the three phase admittance matrix with in and out of phase components as described for $Z$. Note that Eq.~(\ref{p-threephasecomplex}) is the generalization of the Eq.~(\ref{p-complex}) from one to three phases. In the next section, we introduce linear approximation to power flows. It is done first for the single-phase case and then extended to the case of the  three phase unbalanced grids.

\section{Linear Power flow Models}
\label{sec:linearized}
In this section we discuss a linear coupled approximate model (LC-PF3) for three phase unbalanced power flow. We show that it generalizes the linear power flow model (LC-PF) that was discussed in our previous work \cite{dekatcns} for the case of the single-phase network. 

\textbf{Linear Coupled Power Flow (LC-PF) model} \cite{dekatcns,bolognani2013identification}:
In this model, the single-phase PF Eq.~(\ref{p-complex}) is linearized jointly over the phase angle difference between neighboring buses $(\theta_i - \theta_j)$ and deviations of the voltage magnitude $(v_i - 1)$ from the reference voltage of $1$ p.u., both of which are considered to be small. We arrive at the following set of Linear-Coupled (LC) equations:
\begin{align}
p_i&=\underset{j:(ij)\in{\cal E}}{\sum}\left(r_{ij}(v_i-v_j)+ x_{ij}(\theta_i-\theta_j)\right)/\left({x_{ij}^2+r_{ij}^2}\right),
\label{LC-PF_p}\\
q_i&=\underset{j:(ij)\in{\cal E}}{\sum}\left(x_{ij}(v_i-v_j) -r_{ij}(\theta_i-\theta_j)\right)/\left({x_{ij}^2+r_{ij}^2}\right)
\label{LC-PF_q}
\end{align}
One can conveniently express the linear equations of the LC-PF model in the matrix form
\begin{align}
P = p+\tilde{i}q = M^T[Z^*]^{-1}M(v - \tilde{i}\theta),\label{LC-PF}
\end{align}
where $[Z^*]$ is a diagonal matrix where the nonzero elements are complex conjugates of the respective line impedances  and  $M$ is the edge to node incidence matrix for $\cal G$: every edge $(ij) \in {\cal E}$ is represented by a row $M_{ij}= (e_i^T -e_j^T)$, where $e_i$ is the standard basis vector associated with the vertex $i$.
Note that in deriving Eq.~(\ref{LC-PF}) we ignore losses of both active and reactive powers in lines thus getting conservation of the net active and reactive powers.
To make Eq.~(\ref{LC-PF}) invertible one fixes the voltage magnitude to unity and phase to zero at the reference/slack/sub-station bus.
Then injections at the reference bus are equal to the negative sum of injections at all other buses. With these (standard) manipulations we effectively remove the reference bus from the system and without a loss of generality, measure voltage magnitudes and phases at other buses relative to that at the reference bus. Removing entries corresponding to the reference bus from matrix $M$ and vectors $p,q,v, \theta$, we arrive at an invertible full-rank system resulting in
\begin{align}
V^1  = v - \tilde{i}\theta = M^{-1}[Z^*]{M^{-1}}^T(p +\tilde{i}q)\label{reducedsingle}
\end{align}
where $V^1 = v - \tilde{i}\theta$. Next, we describe linear approximations of the three phase power flow.

\textbf{Linear Coupled Three Phase Power Flow (LC-PF3)}: Consider the three phase PF described in Eq.~(\ref{p-threephasecomplex}) with reference phase angle $\hat{\theta}_{ref}$ in Eq.~(\ref{theta_ref}). Here, we consider small deviations in voltage magnitude and phase angle from nominal at each bus and small angle difference between neighboring buses. In three phases, our assumption can be stated as follows
\begin{align}
&\forall i\in{\cal V}:\|\hat{v}_i -\textbf{1}\|= \|[
(v^a_i-1)~(v^b_i-1)~(v^c_i-1)]^T\| \ll 1\nonumber\\
&~~~~~\theta^{\alpha}_i-\theta^{\beta}_i \approx \theta_{ref}^{\alpha\beta} =\theta^{\alpha}_{ref} -\theta^{\beta}_{ref}~\forall \alpha,\beta \in \{a,b,c\}\label{anglediff}\\
&\forall (ij) \in{\cal E}: \|\hat{\theta}_i- \hat{\theta}_j\| \ll 1 
\end{align}
Eq.~(\ref{anglediff}) states that for each node, the angles at different phases are roughly separated by the same amount as the reference phase angles (i.e., by $2\pi/3$). Under these small deviation assumptions, we approximate PF Eq.~(\ref{p-threephasecomplex}) at each phase for node $i\in {\cal V}$ as follows:
\begin{align}
 P^{\alpha}_i&=\smashoperator[l]{\sum_{j:(i,j)\in{\cal E}}}~\smashoperator[r]{\sum_{\beta \in \{a, b,c\}}} v^{\alpha}_i \left(v^{\beta}_i e^{\tilde{i}(\theta^{\alpha}_i -\theta^{\beta}_i)}-v^{\beta}_j e^{\tilde{i}(\theta^{\alpha}_i -\theta^{\beta}_j)}\right)Y^{\alpha\beta}_{ij}\nonumber\\
\Rightarrow P^{\alpha}_i&=\smashoperator[lr]{\sum_{\beta \in \{a, b,c\}}}e^{\tilde{i}\theta_{ref}^{\alpha\beta}}\smashoperator[lr]{\sum_{j:(i,j)\in{\cal E}}}
\left((v^{\beta}_i-v^{\beta}_j) - \tilde{i} (\theta^{\beta}_i-\theta^{\beta}_j)\right)Y^{\alpha\beta}_{ij} \label{LC-PF3}
\end{align}
where Eq.~(\ref{LC-PF3}) is derived similar to Eqs.~(\ref{LC-PF_p},\ref{LC-PF_q}) by ignoring second order terms in voltage magnitude and phase angle differences. The Linear Coupled Power Flow model in three phases (LC-PF3) is given by Eq.~(\ref{LC-PF3}). Note that LC-PF3 equations reduce to the LC-PF Eqs.~(\ref{LC-PF_p},\ref{LC-PF_q}) if the number of phases is limited to one. 
We collect nodal voltage magnitudes and angles for each phase into vectors, and line admittances for each pair of phases into diagonal matrices to express LC-PF3 as a linear equation, similar to the single-phase LC-PF case in Eq.~(\ref{LC-PF})
\begin{align}
P^{\alpha} = p^{\alpha} +\tilde{i}q^{\alpha} = \smashoperator[lr]{\sum_{\beta \in \{a, b,c\}}}e^{\tilde{i}\theta_{ref}^{\alpha\beta}}M^T[Y^{\alpha\beta}]M(v^{\beta}-\tilde{i}\theta^{\beta}).
\end{align}
Combining the expressions for power injections in all three phases, we arrive at
\begin{subequations}
\footnotesize
\begin{align}
& \begin{bmatrix} P^{a}\\ e^{-\tilde{i}2\pi/3}P^{b}\\e^{\tilde{i}2\pi/3}P^{c}\end{bmatrix} = \begin{bmatrix}M~ 0~ 0\\0~ M~ 0\\0~ 0~ M\end{bmatrix}^T\begin{bmatrix}Y^{aa} &Y^{ab} &Y^{ac}\\Y^{ab} &Y^{bb} &Y^{bc}\\Y^{ac} &Y^{bc} &Y^{cc}\end{bmatrix} \begin{bmatrix}M~ 0~ 0\\0~ M~ 0\\0~ 0~ M\end{bmatrix}\begin{bmatrix}v^{a}-\tilde{i}\theta^{a}\\ e^{-\tilde{i}2\pi/3}(v^{b} - \tilde{i}\theta^{b})\\ e^{\tilde{i}2\pi/3}(v^{c}-\tilde{i}\theta^{c})\end{bmatrix}\nonumber\\
&\Rightarrow P^\dag = {M^\dag}^TY^\dag M^\dag V^\dag \label{forinv}\\
&\text{where~} P^\dag = \begin{bmatrix} P^{a}\\ e^{-\tilde{i}2\pi/3}P^{b}\\e^{\tilde{i}2\pi/3}P^{c}\end{bmatrix} , V^\dag = \begin{bmatrix}v^{a}-\tilde{i}\theta^{a}\\ e^{-\tilde{i}2\pi/3}(v^{b} - \tilde{i}\theta^{b})\\ e^{\tilde{i}2\pi/3}(v^{c}-\tilde{i}\theta^{c})\end{bmatrix}, M^\dag = diag(M,M,M).\nonumber
\end{align}
\end{subequations}
Here $Y^\dag$ is a square block matrix where every block, $Y^{\alpha\beta}$, is a diagonal matrix constructed from admittances  of the respective pairs $(\alpha,\beta)$. We also present in the supplementary material, comparison of voltages generated by $LC-PF3$ with those generated by non-linear three-phase unbalanced flows. 
As LC-PF3 is lossless, injections at the reference bus are given by the negative sum of injections at all other nodes. Therefore, removing entries corresponding to the reference bus for each phase in $M^\dag, p^\dag, q^\dag, v^\dag$ and $\theta\dag$, we invert Eq.~(\ref{forinv}) and express the three phase voltages in terms of the three phase injections in the reduced system as follows:
\begin{align}
V^\dag = {M^\dag}^{-1}{Z^\dag}^H{M^{\dag-1}}^TP^\dag \label{reducedthree}
\end{align}
where the structure of ${Z^\dag}^H={Y^\dag}^{-1}$ is provided in Theorem \ref{inverse} in the supplementary material.
Note that both reduced LC-PF Eq.~(\ref{reducedsingle}) and LC-PF3 Eq.~(\ref{reducedthree}) are defined over general grids (possibly loopy). 
The LC-PF3 model over  a radial grid appears equivalent to the lossless approximation of the three phase DistFlow Model. \cite{lowlinear} shows a different derivation of the LC-PF3 which ignores components corresponding to line losses in the power flow equations. Aside from being lossless, another characteristic of the aforementioned models (LC-PF, LC-PF3) is that they represent nodal complex voltages as an invertible function of nodal injections at the non-reference buses. This property is fundamental to determining the graphical model of nodal voltages discussed in the following section. 

\section{Graphical Model from Power Flows}
\label{sec:graphicalmodel}
\begin{figure}[!ht]
\centering
\hfill
\subfigure[]{\includegraphics[width=0.14\textwidth]{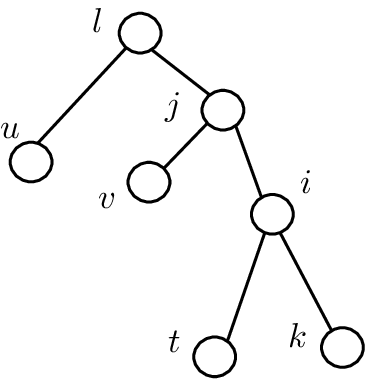}\label{fig:junction1}}\hfill
\subfigure[]{\includegraphics[width=0.14\textwidth]{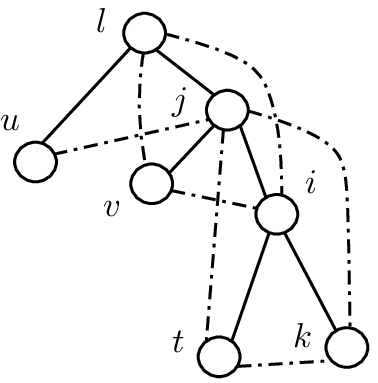}\label{fig:junction2}}\hfill
\vspace{-.25cm}
\caption{(a) Load nodes and edges in distribution tree ${\cal T}$. (b) GM for LC-PF or LC-PF3 in $\cal T$. Dotted lines represent two hop neighbors.}
\label{fig:junction}
\vspace{-3mm}
\end{figure}

We now describe the probability distribution of nodal voltages in the distribution tree-grid ${\cal T}$ (see Fig.~\ref{fig:junction1}) considered under LC-PF model or LC-PF3 model. First we make the following assumptions regarding statistics of the power injections in the distribution tree-grid:

\textbf{Assumption 1}: Loads/injections at all non-substations nodes are modeled as $PQ$ nodes, i.e. for any instance $P$ and $Q$ at a node is kept constant. Loads at the nodes are assumed generated from a probability distribution modeling an exogenous process. Loads at different nodes are statistically independent.

Note that the latter, most important, part of the Assumption 1 concerning the independence is a formalization of the observation that consumers/producers act, e.g. switching on/off their devices, independently at short intervals. Similar assumptions of independence are reported in the literature \cite{dekatcns,saverio}. Note that Assumption $1$ does not require that $P$ and $Q$ components at the same node are uncorrelated. In particular, active and reactive injections at the same node are allowed to be dependent. If each individual load/injection is by itself an accumulation/aggregation of many random processes one would expect, according to the law of large numbers, that fluctuations of the load is well modeled by a Gaussian process \cite{duehee,yury}.

Under Assumption $1$, the continuous random vector of injections at the non-substation nodes within the tree-grap ${\cal T}$, $P$ (under LC-PF Eq.~(\ref{LC-PF})) and $P^\dag$ (under LC-PF3 Eq.~(\ref{forinv})) are described by the following Probability Distribution Functions (PDF):
\begin{align}
{\cal P}(P)=\prod_{i\in {\cal V}}{\cal P}_i(P_i) \quad {\cal P}(P^\dag)=\prod_{i\in {\cal V}}{\cal P}_i(\hat{P}_i) \label{GM_P}
\end{align}
where ${\cal P}_i(P_i)$ and ${\cal P}_i(\hat{P}_i)$ are the p.d.f.s for injection at node $i$ in single phase and three phases respectively. Note that $P^\dag$ and $\hat{P}$ contain the exact information as the relation between them is governed by a phase specific constant rotation.
Using the invertible relations between voltages and injections in the LC-PF and LC-PF3 models, one arrives at the following PDF of the complex voltage vector $V^1$, $V^\dag$ for the non-substation nodes
\begin{align}
{\cal P}(V^1) = \frac{1}{|J_P(V^1)|}{\cal P}(P),~~(P,V^1) \text{~satisfy Eq.}~(\ref{reducedsingle})  \label{GM_V1}\\
{\cal P}(V^\dag) = \frac{1}{|J_P(V^\dag)|}{\cal P}(P^\dag),~~(P^\dag,V^\dag) \text{~satisfy Eq.}~(\ref{reducedthree}), \label{GM_V3}
\end{align}
where $|J_P(V^1)|$ and $|J_P(V^\dag)|$ represent the determinants of the Jacobian matrices for the invertible linear transformation from injections to voltages in the LC-PF and LC-PF3 models respectively. Note that as the transformation is linear, i.e. the Jacobian determinants are constant. We now describe the Graphical Model (GM) representation of the probability distribution for nodal voltages that we use below (in the following section) for topology estimation.

\textbf{Graphical Model}: A $n$ dimensional random vector $X = [X_1, X_2,..X_n]^T$ is described by an undirected graphical model ${\cal GM}$ \cite{wainwright2008graphical} with node set ${\cal V_{GM}}=\{1,\cdots,n\}$ and edge set ${\cal E_{GM}}$ representing conditional dependence: edge $(ij) \in {\cal E_{GM}}$ if and only if $\forall C \subset {\cal V_{GM}} - \{i,j\}, {\cal P}(X_i|X_j,X_C) \neq {\cal P}(X_i|X_C)$. Here $X_C$ represents  random variables corresponding to nodes in the set $C$. Stated differently, the set of neighbors of node $i$ is represented by random variables that are conditionally dependent. It follows from this definition \cite{wainwright2008graphical} that if deletion of a set of nodes $C$ separates the graphical model ${\cal GM}$ into two disjoint sets $A$ and $B$, then each node in $A$ is conditionally independent of a node in $B$ given all nodes in $C$.

We now discuss the LC-PF and LC-PF3 models over the distribution tree-graph $\cal T$ and analyze the structure of the respective GMs of nodal voltages. Note that each node in the GM for single phase voltages represents two scalar variables, the voltage magnitude and phase at the corresponding node. Similarly, each node in the three phase GM corresponds to the complex voltage at each node in three phases (six scalar variables).
Consider LC-PF model. For tree ${\cal T}= \{{\cal V}, {\cal E}\}$, PDF of the single phase voltages 
is given by Eq.~(\ref{GM_V1}). Using Eq.~(\ref{LC-PF},\ref{GM_P}), one derives
\begin{align}
{\cal P}(V^1) &= \frac{1}{|J_P(V^1)|}\prod_{i\in{\cal V}}{\cal P}_i(P_i)\nonumber\\
&=\frac{1}{|J_P(V^1)|}\prod_{i\in{\cal V}}{\cal P}_i\left(\smashoperator[r]{\sum_{j:(ij)\in{\cal E}}}(V^1_a - V^1_b)/Z_{ij}^*\right)\label{GM_long1}
\end{align}
where the Jacobian/determinant is constant. Note that each term under the product sign at the right hand side of Eq.~(\ref{GM_long1}) includes voltages corresponding to a node and all its neighbors in $\cal T$. Consider two nodes $i$ and $j$ in $\cal V$. If $i$ and $j$ are neighbors then terms ${\cal P}_i(P_i),{\cal P}_j(P_j)$ include voltages at both $i$ and $j$. Similarly, if $i$ and $j$ are two hops away and share a common neighbor $k$, voltages at $i$ and $j$ appear in ${\cal P}_k(P_k)$. However for $i$ and $j$ that are three or more hops away, no term includes voltages at $i$ and voltages at $j$. Thus the PDF ${\cal P}(V^1)$ can be product separated into terms containing only $i$ or $j$, but not both. This implies that voltages at two nodes $i$ and $j$ are conditionally independent given voltages at all other nodes if and only if the distance between them in $\cal T$ is greater than $2$ hops.

Next consider the ${\cal GM}_3$ for the PDF of the three phase voltages (Eq.~(\ref{GM_V3})). Using LC-PF3 Eq.~(\ref{forinv}) with Eq.~(\ref{GM_P}), the PDF of nodal voltages can be expanded as
\begin{align*}
\footnotesize
&{\cal P}(V^\dag) = \frac{1}{|J_P(V^\dag)|}\prod_{i\in {\cal V}}{\cal P}_i(\hat{P}_i) \\
=\frac{1}{|J_P(V^\dag)|}&\prod_{i\in{\cal V}}{\cal P}_i\left(\smashoperator[r]{\sum_{j:(ij)\in{\cal E}}}~
\begin{bmatrix}
&\smashoperator[r]{\sum_{\beta \in \{a, b,c\}}}e^{\tilde{i}\theta_{ref}^{a\beta}}(V^{a}_i - V^{\beta}_j)Y^{a\beta}_{ij}\\
&\smashoperator[r]{\sum_{\beta \in \{a, b,c\}}}e^{\tilde{i}\theta_{ref}^{b\beta}}(V^{b}_i - V^{\beta}_j)Y^{b\beta}_{ij}\\
&\smashoperator[r]{\sum_{\beta \in \{a, b,c\}}}e^{\tilde{i}\theta_{ref}^{c\beta}}(V^{c}_i - V^{\beta}_j)Y^{c\beta}_{ij}
\end{bmatrix}
\right).
\end{align*}
Using the same analysis as that for the single phase case, one observes that the PDF of nodal voltages in LC-PF3 has a similar feature with voltages at nodes greater than two hops being conditionally independent. Using the aforementioned definition of conditional independence and GM structure, we arrive at the following lemma.
\begin{lemma} \label{lemma1}
Graphical models ${\cal GM}_1$ for the PDF of single phase voltages (Eq.~(\ref{GM_V1})) and ${\cal GM}_3$ for the PDF of three phase voltages (Eq.~(\ref{GM_V3})) contains edges between single and two hops nodes in tree-graph $\cal T$.
\end{lemma}
Fig.~\ref{fig:junction2} shows an example construction of a GM correspondent to either of the aforementioned power flow models. Each node in $\cal GM$ represents the single or three phase voltage at its corresponding node in $\cal T$. Note that the GM, unlike the tree-graph $\cal T$ itself, is loopy due to edges between nodes separated by two hops in ${\cal T}$. 

\subsection{The Gaussian Case}
\label{sec:gaussian}
In prior work \cite{dekatcns, saverio, duehee, zhu, yury} loads have been modeled as independent Gaussian random variables. As linear functions of Gaussian random variables are Gaussian random variables too, the distribution of nodal voltages in LC-PF or LC-PF3 under the Gaussian assumption is a multivariate Gaussian. For multi-variate Gaussian distribution, two properties are particularly useful \cite{wainwright2008graphical}:
\begin{itemize}[leftmargin = *]
\item The structure of the GM is given by the non-zero off-diagonal terms in the inverse covariance matrix of the random vector.
\item Variables $i$ and $j$ are conditionally independent given variables in set $C$ if the \textit{conditional covariance} of $i$ and $j$, given set $C$, is zero.
\end{itemize}
The first property can be used to validate Lemma \ref{lemma1} for the Gaussian GM as shown in the supplementary material. In the next section, we use the specific structure of the GM described in this section to develop our topology learning algorithm for voltage measurements, in particular arising from the Gaussian distributions.

\section{Topology Learning using Voltage Conditional Independence}
\label{sec:learning}
Consider the tree-grid $\cal T$ with PDFs of nodal voltages ${\cal P}(V^1)$  and ${\cal P}(V^\dag)$ correspondent, respectively, to LC-PF and LC-PF3. Let the corresponding GM for voltages be  ${\cal GM}_1$ (single phase) and ${\cal GM}_3$ (three phase). As described in the previous section, the GM includes edges between true neighbors and two hop neighbors in $\cal T$. Our topology learning algorithm is based on `separability' properties of the GMs that are only satisfied for edges corresponding to true neighbors in $\cal T$. To learn the topology without ambiguity, we make the following mild assumption on the structure of the operational tree-grid $\cal T$.

\textbf{Assumption 2}: The depth (length of the longest path) of the tree-grid $\cal T$ (excluding the substation node) is greater than three.

Note that Assumption $2$ is satisfied when $\cal T$ includes at least \textit{two non-leaf nodes that are two or more hops away}. This is thus not restrictive for the majority of distribution grids (real world and test cases) that have long paths. Under Assumption $2$, the next theorem lists conditional independence properties in voltage distributions that distinguish true edges. We use the following notation for conditional independence of random variables $X,Y$ given variables $A,B$:
\begin{align}
X\indep Y|(A,B) ~~\equiv~~ {\cal P}(X,Y|A,B) = {\cal P}(X|A,B){\cal P}(Y|A,B)
\end{align}

\begin{theorem}\label{condindnonleaf}
$(ij)$ is an operational edge between non-leaf nodes $i$ and $j$ in  $\cal T$ if and only if there exists distinct nodes $k$ and $l$ such that $V^1_k\indep V^1_l|(V^1_i,V^1_j)$ (for single phase) and $V^\dag_k\indep V^\dag_l|(V^\dag_i,V^\dag_j)$ (for three phase).
\end{theorem}
The proof is given in the supplementary material. Theorem \ref{condindnonleaf} enables detection of edges between all non-leaf nodes using their voltage measurements. Let ${\cal T}_{nl}$ (see Fig.~\ref{fig:reconstruct1}) comprise of connections between all non-leaf nodes ${\cal V}_{nl}$ in grid $\cal T$ (estimated using Theorem \ref{condindnonleaf}). Let ${\cal V}^1_{nl}$ be the set of nodes of degree $1$ in ${\cal T}_{nl}$ (Eg. node $i$). Note that ${\cal V}^1_{nl}$ consists of non-leaf nodes in grid $\cal T$ that are neighbors of only one non-leaf node. Then the next theorem provides results helping to determine true parent of each leaf node in $\cal T$.
\begin{theorem}\label{condindleaf}
Let ${\cal T}_{nl}$ be sub-graph of  $\cal T$ with non-leaf nodes and respective edges removed.
Let ${\cal V}^1_{nl}$ be the set of nodes of degree $1$ in ${\cal T}_{nl}$. Then the following statement holds:
\begin{enumerate}[leftmargin=*]
\item Node $i \in {\cal V}^1_{nl}$ is the parent of leaf node $k$ in $\cal T$ if and only if \textbf{for any} nodes $j,l \in {\cal V}_{nl}$ with edges $(ij), (jl)$, $V^1_k\indep V^1_l|(V^1_i,V^1_j)$ (for single phase) and $V^\dag_k\indep V^\dag_l|(V^\dag_i,V^\dag_j)$ (for three phase).
\item Let leaf node $k$'s parent be in set ${\cal V}_{nl} - {\cal V}^1_{nl}$. Then $i$ is the parent of $k$ if and only if \textbf{for all} nodes $j,l \in {\cal T}-\{i,k\}$ with edges $(ij),(jl)$, $V^1_k\indep V^1_l|(V^1_i,V^1_j)$ (for single phase) and $V^\dag_k\indep V^\dag_l|(V^\dag_i,V^\dag_j)$ (for three phase).
\end{enumerate}
\end{theorem}
The proof is presented in the supplementary material. Let us emphasize that the first result in Theorem \ref{condindleaf} identifies connections between leaves and parents that have a single non-leaf neighbor (Eg. parent $i$ in Fig.~\ref{fig:junction}). The remaining leaves are children of nodes with two or more non-leaf neighbors (Eg. non-leaf node $j$ in Fig.~\ref{fig:junction}). Such connections are identified by the second result in the theorem.

Theorems \ref{condindnonleaf} and \ref{condindleaf} imply the topology learning steps in Algorithm $1$. \begin{algorithm}
\caption{Topology Learning for Grid Tree ${\cal T}$}
\textbf{Input:} Complex voltage observations $V^{in}_j = V^1_j$ (single phase) or $V^\dag_j$ (three phase) at all non-substation nodes $j \in {\cal V}$, Permissible edge set $\mathcal{E}_{full}$\\
\textbf{Output:} Operational edge set ${\cal E}$\\
\begin{algorithmic}[1]
\State ${\cal E},{\cal V}_{nl} \gets \emptyset$
\ForAll{$(ij) \in \mathcal{E}_{full}$} \label{nonleaf1}
\If {$\exists k,l \in {\cal V}-\{i,j\}$ s.t. $V^{in}_k\indep V^{in}_l|(V^{in}_i,V^{in}_j)$} \label{combo}
\State ${\cal E} \gets {\cal E} \cup \{(ij)\}$, ${\cal V}_{nl} \gets {\cal V}_{nl} \cup \{i,j\}$
\EndIf
\EndFor\label{nonleaf2}

\State ${\cal T}_{nl} = \{{\cal V}_{nl}, {\cal E}\}$ \label{treenonleaf}
\State ${\cal V}^1_{nl} \gets \{\text{nodes of degree $1$ in~}{\cal T}_{nl}\}, {\cal V}^2_{nl}= {\cal V}_{nl} - {\cal V}^1_{nl}$
\ForAll{$k \in {\cal V} - {\cal V}_{nl}$} \label{leaf1}
\ForAll{$i \in {\cal V}^1_{nl}$, $(ki) in \mathcal{E}_{full}$}
\State Pick $j,l \in {\cal V}_{nl}$ with $(ij),(jl) \in {\cal E}$
\If {$V^{in}_k\indep V^{in}_l|(V^{in}_i,V^{in}_j)$}
\State ${\cal E} \gets {\cal E} \cup \{(ik)\}$, ${\cal V}_{nl} \gets {\cal V}_{nl} \cup \{k\} $
\EndIf
\EndFor
\EndFor \label{leaf2}

\ForAll{$k \in {\cal V} - {\cal V}_{nl}$} \label{leaf3}
\ForAll{$i \in {\cal V}^2_{nl}$, $(ki) \in \mathcal{E}_{full}$}
\If {$V^{in}_k\indep V^{in}_l|(V^{in}_i,V^{in}_j) \forall j,l \in {\cal V}_{nl}$,~$(ij),(jl) \in {\cal E}$}
\State ${\cal E} \gets {\cal E} \cup \{(ik)\}$, ${\cal V}_{nl} \gets {\cal V}_{nl} \cup \{k\} $
\EndIf
\EndFor
\EndFor \label{leaf4}
\State ${\cal T} \gets \{{\cal V}_{nl}, {\cal E}\}$
\end{algorithmic}
\end{algorithm}

\textbf{Execution}:
Our topology learning algorithm proceeds in three steps. First, edges between non-leaf nodes are identified based on Theorem  \ref{condindnonleaf} in Steps (\ref{nonleaf1}-\ref{nonleaf2}) and radial network of non-leaf nodes ${\cal T}_{nl}$ is constructed in Step (\ref{treenonleaf}). Next, leaves in $\cal T$ connected to nodes of degree $1$ (set ${\cal V}^1_{nl}$) in ${\cal T}_{nl}$ are identified using Theorem \ref{condindleaf}(1) in Steps (\ref{leaf1}-\ref{leaf2}). Finally edges between leaves and non-leaf nodes connected to two or more non-leaf nodes (set ${\cal V}^2_{nl}$) are identified using Theorem \ref{condindleaf}(2) in Steps (\ref{leaf3}-\ref{leaf4}). It is worth mentioning that the learning algorithm does not require any other information other than the voltage measurements at the grid nodes. It does not require information of line impedances or statistics of nodal injections. If the set of permissible lines is not available, all node pairs are considered as permissible edges in set $\mathcal{E}_{full}$. 
The steps in reconstruction for the radial grid in Fig.~\ref{fig:junction1} is given in the supplementary material as an example along with the analysis of the computational complexity of the learning algorithm. Furthermore we discuss in detail the computation of conditional independence tests, in particular for Gaussian distributions, in the supplementary material. In the next section we present simulation results of Algorithm $1$ over linearized and non-linear samples in single and three phase systems.
\section{Simulations Results}
\label{sec:simulations}
We test Algorithm $1$ by extracting the operational edge set ${\cal E}$ of tree grid ${\cal T}$ from a loopy original edge set ${\cal E}_{full}$. If ${\cal E}_{full}$ is not available, all node pairs are considered as potential edges.

\begin{figure}[!bt]
\centering\hfill
\subfigure[]{\includegraphics[width=0.18\textwidth,height =.14\textwidth, angle=-90]{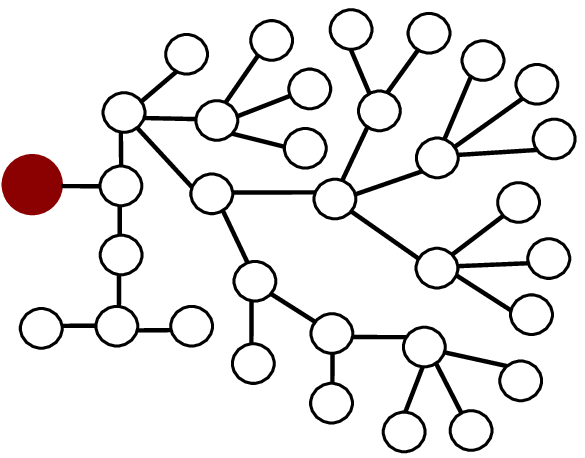}\label{fig:case33}}\hfill
\subfigure[]{\includegraphics[width=0.09\textwidth,height = .06\textwidth,angle=-90]{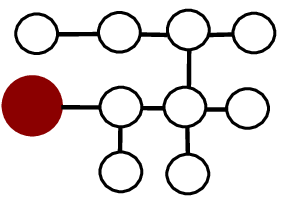}\label{fig:case10}}\hfill
\vspace{-.10cm}
\subfigure[]{\includegraphics[width=0.19\textwidth,height = .14\textwidth,angle=-90]{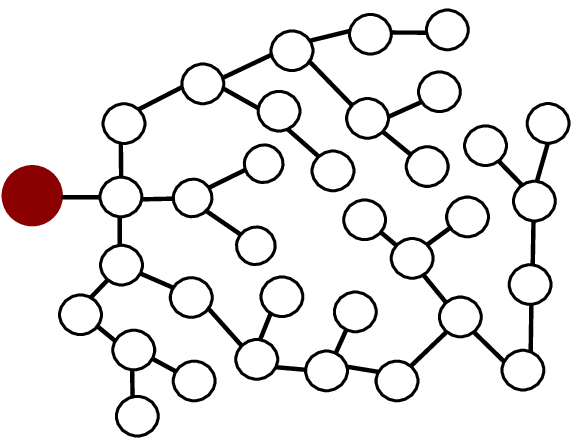}\label{fig:case35}}\hfill
\caption{Layouts of the modified IEEE test distribution grids. The red circles represent substations. (a) Single-phase $33$-bus network \cite{matpower} (b) Three phase $10$-bus network \cite{kersting2001radial} (c) Three phase $35$-bus network \cite{kersting2001radial}}
\end{figure}

\begin{figure}[!bt]
\centering
\includegraphics[width=0.30\textwidth,height = .28\textwidth]{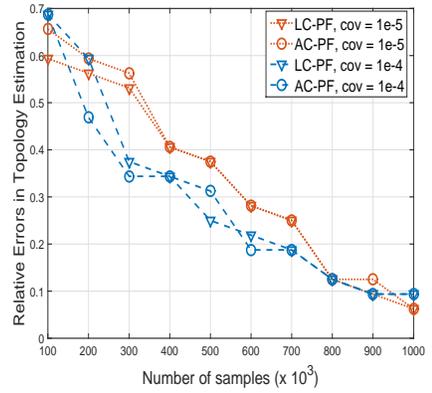}
\caption{Accuracy of topology learning algorithm with increasing number of voltage samples generated by LC-PF and AC-PF for $33$ bus test case in Fig.~\ref{fig:case33}. Injection covariances are taken to be $10^-4$ and $10^-5$. Permissible edge set $\mathcal{E}_{full}$ has $76$ edges}.
\label{fig:adj_errors_33}
\vspace{-3mm}
\end{figure}

\begin{figure}[!ht]
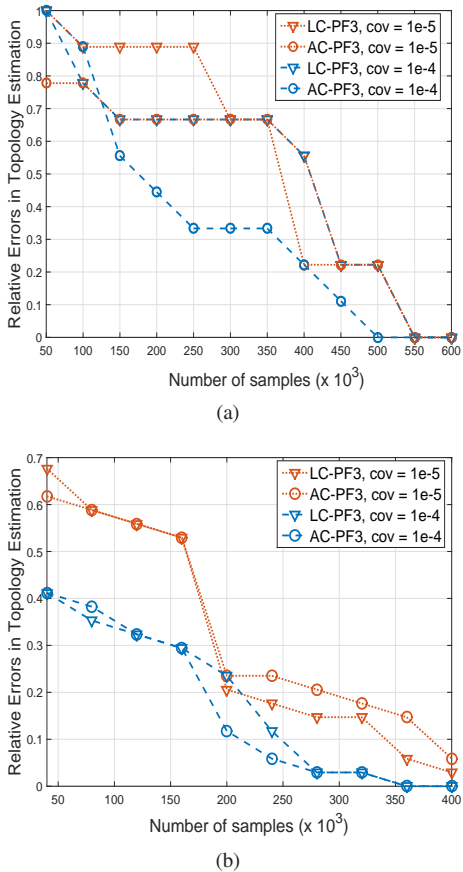

\centering
\subfigure[]{\includegraphics[width=0.33\textwidth,height = .28\textwidth]{adj_errors_10_three.eps}\label{fig:adj_errors_10_full}}\hfill
\subfigure[]{\includegraphics[width=0.33\textwidth,height = .28\textwidth]{adj_errors_35_three.eps}\label{fig:adj_errors_35}}
\caption{Accuracy of topology learning algorithm with increasing number of three phase voltage samples generated by LC-PF3 and AC-PF3 with injection covariances $10^{-4}$ and $10^{-5}$. (a) $10$ bus test case in Fig.~\ref{fig:case10}. All non-substation node pairs are considered as permissible edges. (b) $35$ bus test case in Fig.~\ref{fig:case35}. Permissible edge set $\mathcal{E}_{full}$ has $84$ edges.}
\vspace{-2mm}
\end{figure}

First, we discuss topology estimation using voltage samples generated by single phase non-linear (lossy) AC-PF equations. We consider a radial modification of the $33$-bus system \cite{matpower} shown in Fig.~\ref{fig:case33}. The input set $\mathcal{E}_{full}$ comprises of $76$ edges ($32$ true and $44$ additional edges selected randomly). As before, we consider Gaussian load fluctuations and generate AC-PF samples using Matpower \cite{matpower}.  We show performance of the Algorithm $1$ for different sample sizes and two distinct injection covariance tests in Fig.~\ref{fig:adj_errors_33}. For comparison, we also present the performance of Algorithm $1$ over LC-PF voltage samples generated from the same injection samples as in the AC-PF model. Observe that the performance over AC-PF voltage samples is similar to the one observed in LC-PF and it improves with sample size increase, thus confirming that Algorithm $1$, even though built on the linearization principles, performance empirically well over the data generated from the
non-linear AC power flow models.
Next, we discuss performance of the three phase power flow models $10$ and $35$ bus test cases \cite{linearpf3} modified from IEEE $13$ and $37$ test cases \cite{kersting2001radial}. We consider bus $1$ as reference node in both three phase test networks for our topology learning algorithm as it has degree $1$.  We consider two different Gaussian nodal injection covariances ($10^{-5}$ and $10^{-4}$) in both networks and generate input voltage samples (using LC-PF3 and AC-PF3). For the three phase $10$-bus network, we include all node pairs as permissible edges in $\mathcal{E}_{full}$. The performance of Algorithm $1$ for different samples sizes for this case is depicted in Fig.~\ref{fig:adj_errors_10_full}. For the $35$ bus network, we pick at random $50$ additional edges added to the true edges and input a permissible edge set $\mathcal{E}_{full}$ of size $84$ to Algorithm $1$ along with the three phase voltage samples. The relative errors in topology estimation for different input sample sizes for this network are shown in Fig.~\ref{fig:adj_errors_35}. Note that for either of three phase networks, the errors for non-linear AC power flows are less or comparable to errors observed in  linerized power flow model LC-PF3. Furthermore, the errors decrease with increase in the sample size. As before, we optimize values of the thresholds for $\text{\textbf{cond}}_\text{\textbf{mod}}$\textbf{-test} used in Algorithm $1$ (see supplementary material) using trial and error search that, in practice, can be determined from historical or simulated data.
\section{Conclusion}
\label{sec:conclusion}
In this paper, we develop algorithm which allows to estimate the radial topology of distribution grids. In particular, we derive linearized power flow model in single and unbalanced three phase cases and develop a Graphical Model based learning algorithm that is able to estimate operational topology of the networks from samples of nodal voltages. Our learning algorithm is very general as it does not require information on nodal injection statistics or line parameters. To the best of our knowledge this is the first approach which develops algorithm with guarantees for topology estimation in both balanced (effectively single phase) and unbalanced (three phase) networks. 
We demonstrate empirical efficacy of our algorithm on a number of AC-nonlinear IEEE test cases.

This work has a number of promising future extensions. First, realistic networks may have portions where the three phase layout is split into three single-phase lines of different lengths.  Extension of our algorithm to this case is straightforward.
Second, the linear flow model based topology learning can be used jointly with phase identification and impendance estimation (see \cite{sejunpscc} for an example of the latter in the single phase case). 
Finally, we plan to extend our empirical AC (nonlinear) approach towards establishing rigorous bounds on the errors between linearized and non-liner flow models. 
\bibliography{FIDVR,SmartGrid,voltage,trees,sigproc}

\newpage
\section{Supplementary Material}
\subsection{Inverse of Three-phase }
Note that $Y^\dag$ is a square block matrix where each block is a diagonal matrix. Due to this specific structure, inverse of $Y^\dag$ has a similar block sparse pattern as $Y^\dag$ and the following holds.
\begin{theorem}\label{inverse}
Let $\hat{Y}_{ij}$ and $\hat{Z}_{ij}$ be the three phase admittance and impedance matrices respectively for edge $(ij) \in {\cal E}$, where $\hat{Y}_{ij}^{-1} = \hat{Z}^H_{ij}$. Define $\footnotesize Y^\dag = \begin{bmatrix}Y^{aa} &Y^{ab} &Y^{ac}\\Y^{ab} &Y^{bb} &Y^{bc}\\Y^{ac} &Y^{bc} &Y^{cc}\end{bmatrix}$ where each block is a diagonal matrix with admittances on all lines in $\cal E$ for a phase pair. Define $\footnotesize Z^\dag = \begin{bmatrix}Z^{aa} &Z^{ab} &Z^{ac}\\Z^{ab} &Z^{bb} &Z^{bc}\\Z^{ac} &Z^{bc} &Z^{cc}\end{bmatrix}$ similarly. Then, the inverse of $Y^\dag$ takes the following form
\begin{align}
\footnotesize
{Y^\dag}^{-1} = {Z^\dag}^H =
\begin{bmatrix}&{Z^H}^{aa}~ &{Z^H}^{ab}~ &{Z^H}^{ac}\\&{Z^H}^{ab}~ &{Z^H}^{bb}~ &{Z^H}^{bc}\\&{Z^H}^{ac}~ &{Z^H}^{bc}~ &{Z^H}^{cc}\end{bmatrix}.
\end{align}
\end{theorem}
The proof is omitted. (It reduces to showing that $Y^\dag{Z^\dag}^H$ is equal to an identity matrix.)

\subsection{Validity of linearized three-phase unbalanced power flow model}
\begin{figure}[!ht]
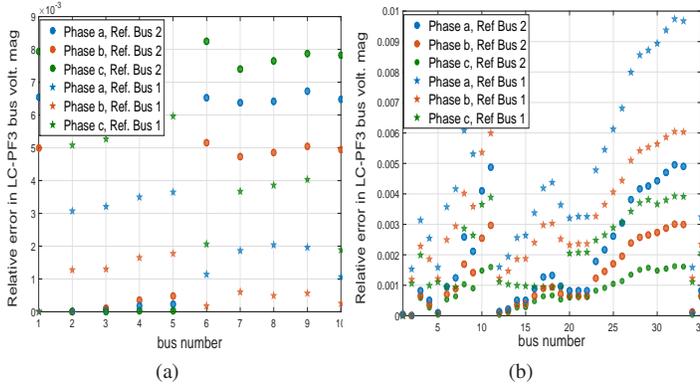

\centering\hfill
\subfigure[]{\includegraphics[width=0.25\textwidth,height = .2528\textwidth]{voltage_10_three.eps}\label{fig:volt_errors_10}}
\subfigure[]{\includegraphics[width=0.26\textwidth,height = .249\textwidth]{voltage_35_three.eps}\label{fig:volt_errors_35}}\hfill
\caption{Accuracy of voltages generated by linear LC-PF3 relative to voltages from non-linear AC-PF3 model for test systems in Fig.~\ref{fig:case10} and \ref{fig:case35} with base load and selection of reference buses ($1$ or $2$).}
\label{fig:volt_errors}
\vspace{-3mm}
\end{figure}

Here we compare three phase voltages generated by linearized model LC-PF3 with that of non-linear three phase power flow model (AC-PF3). We consider three phase $10$ and $35$ bus test cases \cite{linearpf3} that have been modified from IEEE $13$ and $37$ test cases \cite{kersting2001radial}. We modify all nodal loads to be three phase, remove shunts and make all line impedances to be $Y$-connected with three phases. The networks are depicted in Figs.~\ref{fig:case10} and \ref{fig:case35}. Fig.~\ref{fig:volt_errors_10} and \ref{fig:volt_errors_35} show relative errors in bus voltage magnitudes in each phase for LC-PF3 with respect to the true $AC-PF3$ values generated by a conventional back-forward sweep method. The results for each test network include two choices of the reference bus (bus $2$ or bus $1$). Note that the maximum relative error is less than $1\%$ for both networks and any choice of the reference bus. This motivates us to evaluate the performance of topology identification using true three phase voltages generated by AC-PF3 and compare it with LC-PF3.

\subsection{Validation of Lemma \ref{lemma1} under Gaussian Injections}
As mentioned in Section \ref{sec:gaussian}, structure of Gaussian GM is given by the non-zero off-diagonal entries in the inverse covariance matrix. Consider the LC-PF Eq.(~\ref{LC-PF}) where the vector of injection profiles follows an uncorrelated multi-variate Gaussian distribution with diagonal covariance matrices $\Omega_{p}$, $\Omega_{q}$, $\Omega_{pq}$ denoting the variance of active, reactive injections and covariance of active and reactive injections respectively. The covariance matrix of complex voltages $\Omega_{V^1} = \mathbb{E}\left[V^1 -\mathbb{E}[V^1]\right] \left[V^1-\mathbb{E}[V^1]\right]^{T*}$ satisfies
$$\Omega_{V^1}(i,j) = \sum_{k}H^{-1}_{1/Z^*}(i,k)(\Omega_{p}(k,k)+\Omega_{q}(k,k))H^{-1}_{1/Z}(k,j)$$
where $H_{1/Z^*} = M^T[Z^*]^{-1}M$ is the reduced weighted Laplacian matrix for tree ${\cal T}$, with weight for each edge $(ij)$ given by $1/Z^*_{ij}$. One arrives at
\begin{align}
\Omega^{-1}_{V^1}(i,j)=\footnotesize{\begin{cases}&H_{1/Z}(i,j)H_{1/Z^*}(j,j)(\Omega_{p}(j,j)+\Omega_{q}(j,j))^{-1}+\nonumber\\
 &H_{1/Z}(i,i)H_{1/Z^*}(i,j)(\Omega_{p}(i,i)+\Omega_{q}(i,i))^{-1} ~~\text{if $(ij) \in {\cal E}$},\\
&H_{1/Z}(i,k)H_{1/Z^*}(j,k)(\Omega_{p}(k,k)+\Omega_{q}(k,k))^{-1}~~\text{if $(ik),(jk) \in {\cal E}$},\\
&0 ~~\text{otherwise}
\end{cases}} \label{treeinv}
\end{align}
We observe that the GM contains edges between nodes that are separated by less than three hops in $\cal T$, as proposed above. 
The inverse covariance matrix for voltages in LC-PF3 model under Gaussian injections can be derived in a similar way.
\subsection{Proof of Theorem \ref{condindnonleaf}}
\textbf{Note:} ${\cal GM}_1$ (single phase) and ${\cal GM}_3$ (three phase) includes edges between node pairs in $\cal T$ that are one or two hops away (see Lemma \ref{lemma1}). Below we present the proof for ${\cal GM}_1$ only as its extension to ${\cal GM}_3$ is straightforward.

For the \textit{if} part, consider nodes $i,j$ such that voltages at $k,l$ are conditionally independent given voltages at $i,j$. This means that removing nodes $i,j$ from ${\cal GM}_1$ separates nodes $k,l$ into disjoint groups. We prove that $(ij)$ is an edge between non-leaf nodes $i,j$ by contradiction. Let ${\cal P}_{kl}$ be the unique path in $\cal T$ between nodes $k,l$. For separability of $k,l$, at least one of nodes $i,j$ is included in ${\cal P}_{kl}$. Let node $j$ be excluded and ${\cal P}_{kl}: k-\pi_{i-1}-i-\pi_{i+1}-..-l$. Edge $(\pi_{i-1}\pi_{i+1})$ exists in the GM as they are two hop neighbors. Thus removing $i,j$ does not disconnect $k,l$ when only one of $i,j$ is included in ${\cal P}_{kl}$. Finally, consider ${\cal P}_{kl}:k-\pi_{i-1}-i-\pi_{i+1}-\pi_{j-1}-j-\pi_{j+1}-..-l$, such that there is at least one node between $i$ and $j$. Due to edges between two hop neighbors in the GM, removing $i,j$ does not disconnect $k,l$. Note that as leaf nodes are not part of any path between two other distinct nodes, $i,j$ are both non-leaf nodes. Hence $(ij)$ has to be an edge between non-leaf nodes in $\cal T$, by contradiction. For the \textit{only if} part, consider non-leaf neighbors $i,j$ in $\cal T$. There exits neighbor $k$ of $i$ and neighbor $l$ of $j$ in ${\cal T}$ as shown in Fig.~\ref{fig:junction2} with corresponding edges $(ki),(kj),(il),(lj),(ij)$ in graphical model ${\cal GM}_1$. Every path from $k$ to $l$ in ${\cal GM}_1$ includes an edge in $\{(ki),(kj)\}$ and  $\{(il),(jl)\}$. Removing nodes $i,j$ thus disconnects nodes $k,l$ in ${\cal GM}_1$ and makes voltages at $k,l$ conditionally independent.
\subsection{Proof of Theorem \ref{condindleaf}}
\begin{enumerate}[leftmargin=*]
\item $i \in {\cal V}^1_{nl}$ is connected to one other non-leaf node. By Assumption $2$, there exist non-leaf nodes $j,l$ such that $(ij),(jl)$ are edges in $\cal T$. If $k$ is connected to $i$, using the steps in the proof of Theorem \ref{condindnonleaf}, voltages at $k,l$ are conditionally independent given voltages at $i,j$. For the converse let $i^*\neq i$ be the true parent of $k$. Then, path ${\cal P}_{kl}$ from $k$ to $l$ in $\cal T$ does not include node $i$ as $i$ is connected to only one non-leaf node. If ${\cal P}_{kl}$ does not include $j$ then $k,l$  are not disconnected in  ${\cal GM}_1$ or ${\cal GM}_3$ after removing nodes $i,j$. Otherwise, if ${\cal P}_{kl}: k-i^*-..-j-l$, there exists a path from $k$ to $l$ in the GM containing the two hop neighbors of $j$ after removing $i,j$. Therefore, the relation does not hold if $i$ is not the parent of $k$.
\item Let non-leaf node $i^* \in  {\cal V}_{nl} - {\cal V}^1_{nl}$ be the true parent of leaf node $k$. If edges $(i^*j),(jl)$ exist then using similar argument as Theorem \ref{condindnonleaf}, the conditional independence relation holds. For the converse, consider $i \in  {\cal V}_{nl} - {\cal V}^1_{nl}, i \neq i^*$. Consider path ${\cal P}_{ik}: i-\pi_1-\pi_2-..i^*-k$ in $\cal T$. If $k,i$ are separated by more than two hops, voltages at $k,\pi_2$ are not conditionally independent given voltages at nodes $i,\pi_1$. Next consider ${\cal P}_{ik}: i-i^*-k$ has exactly two hops. As $i^* \in  {\cal V}_{nl} - {\cal V}^1_{nl}$, $i^*$ has a non-leaf neighbor $r \not\in {\cal P}_{ik}$. The assumption about $k,r$ violates the relation as edge $(kr)$ belongs to the GM. Therefore the conditional independence relation is satisfied only by the true parent of $k$ in ${\cal V}_{nl} - {\cal V}^1_{nl}$.
\end{enumerate}

\subsection{Reconstruction Steps for Algorithm $1$}
\begin{figure}[!bt]
\centering
\hfill
\subfigure[]{\includegraphics[width=0.14\textwidth,height =.14\textwidth]{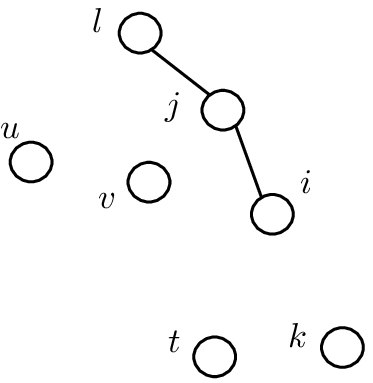}\label{fig:reconstruct1}}\hfill
\subfigure[]{\includegraphics[width=0.14\textwidth,height =.14\textwidth]{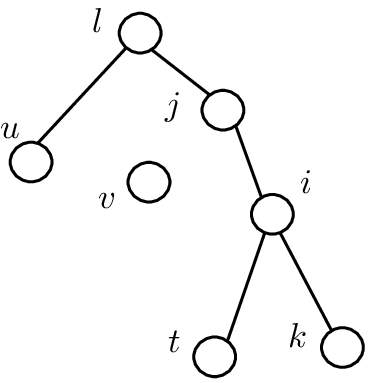}\label{fig:reconstruct2}}\hfill
\subfigure[]{\includegraphics[width=0.14\textwidth,height =.14\textwidth]{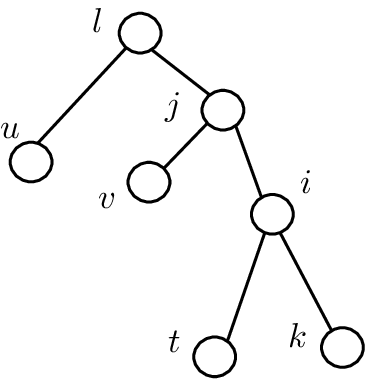}\label{fig:reconstruct3}}\hfill
\vspace{-.25cm}
\caption{Learning steps in Algorithm $1$ for radial grid in Fig.~\ref{fig:junction1} (a) Learning edges between non-leaf nodes (b) Determining children of nodes that are neighbors of one non-leaf node (c) Determining children of nodes that have more than one non-leaf neighbors.}
\label{fig:reconstruct}
\vspace{-3mm}
\end{figure}

Algorithm $1$ when applied to voltages of nodes at Fig.~\ref{fig:junction1} will proceed in the steps listed in Fig.~\ref{fig:reconstruct}.

\subsection{Computational Complexity of Algorithm $1$}
Edge detection in Algorithm $1$ depends on the conditional independence tests. Each test is conducted over voltages at four nodes only (thus quartet). Unlike in the case of a general GM, the computational complexity $C$ in each test is thus \emph{independent of the size of the network $N$}. Identifying the edge between a pair of non-leaf nodes $i,j$ requires $O(N^2)$ tests in the worst case as all combinations of $k,l$ in Step (\ref{combo}) are considered. Therefore, total complexity of identifying the network of non-leaf nodes is $O(N^4)$. Determining the nodes of degree $1$ in ${\cal T}_{nl}$ has complexity $O(N)$. Edge between leaf $k$ and degree one node $i$ in ${\cal T}_{nl}$ can be verified by conditional independent test with a single neighbor and two hop neighbor of $i$. Therefore, complexity of the edge detection of nodes (in ${\cal T}_{nl}$) of degree one and leaves has complexity $O(N^2)$ as number of leaves in $\cal T$ and ${\cal T}_{nl}$ can be $O(N)$. Finally, all combinations of neighbors and two hop neighbors are needed to verify leaves in ${\cal V}^2_{nl}= {\cal V}_{nl} - {\cal V}^1_{nl}$. Steps (\ref{leaf3}-\ref{leaf4}) thus have complexity $O(N^4)$. The overall worst-case complexity of the algorithm is $O(N^4C)$ where $C$ is independent of the network size $N$. Note that we do not assume any prior information of the number of edges or max-degree of a node. For example, if a set of permissible edges $\mathcal{E}_{full}$ is given, then edge detection tests can be restricted to that set. The complexity will then reduce to $O(N^2|\mathcal{E}_{full}|C)$.

\subsection{Conditional Independence Tests}
\label{sec:conditional}
Algorithm $1$ performs the edge detection test for each edge by verifying if the complex voltages at nodes $k,l$ are conditionally independent given voltages at two other nodes $i,j$. To reduce complexity, we check for conditional independence of voltage magnitudes in one phase at $k,l$ given complex voltages at $i,j$. Note that complex voltage at a node in the single phase, LC-PF case consists of two scalars (voltage magnitude and phase angle), and correspondingly of $6$ scalars in three phase, LC-PF3, case. The total number of scalar variables per conditional independence test is $6 (= 2 + 2\times2)$ in the LC-PF case and $14 (= 2+6\times2)$ in the LC-PF3 case.

\textbf{General Voltage Distributions:} As voltage measurements are continuous random variables, testing their conditional independence for general distributions is a non-trivial task. Among non-parametric tests for conditional independence, distances between estimated conditional densities \cite{su2008nonparametric} or between characteristic functions \cite{su2003consistent} have been proposed. One can also bin the domain of continuous values and use discrete valued conditional independence test \cite{margaritis2005distribution}. Another line of work \cite{fukumizu2004dimensionality,gretton2007kernel,zhang2012kernel} focuses on kernel-based conditional independence tests. In these schemes, conditional independence is characterized using vanishing Hilbert-Schmidt norm of covariance operators in Reproducing Kernel Hilbert Spaces (RKHS).

\textbf{Gaussian Voltage Distribution:}
We discuss the special case for Gaussian nodal voltages in detail as they are used in our numerical simulations. Voltages are Gaussian distributed if the loads/injections are Gaussian and relations between loads and voltages are linearwithin LC-PF and LC-PF3.
As noted in Section \ref{sec:gaussian}, conditional independence of the Gaussian random variables is equivalent to vanishing conditional covariance. Consider the following real covariance matrices for LC-PF and LC-PF3.
\begin{align*}
\Sigma^{in}_{kl,ij} &=\mathbb{E}\left[X^{in}_{kl,ij}-\mathbb{E}[X^{in}_{kl,ij}]\right]\left[X^{in}_{kl,ij}-\mathbb{E}[X^{in}_{kl,ij}]\right]^T, ~~\text{where} \nonumber\\
X^{in}_{kl,ij} &= \begin{cases*}[v_k~v_l~v_i~\theta_i~v_j~\theta_j]^T  \text{~in LC-PF},\\   
[v^a_k v^a_l~Re(V^\dag_i)~Im(V^\dag_i)~Re(V^\dag_i)~Im(V^\dag_j)]^T\text{~in LC-PF3}\end{cases*}\nonumber
\end{align*}
where $Re(V^\dag_i)$ and $Im(V^\dag_i)$ refer to the real and imaginary parts of complex vector $V^\dag_i$. Thus, voltages at $k,l$ are conditionally independent given voltages at $i,j$ if the following hold for the $(1,2)^{th}$ entry in the inverse
\begin{align}
   \left[\Sigma^{in}_{kl,ij}\right]^{-1}_{(1,2)} = 0
\end{align}
Note that in LC-PF, $\Sigma^{in}_{kl,ij}$ and its inverse are of size $6\times6$, while $\Sigma^{in}_{kl,ij}$ in LC-PF3 is of size $14\times14$. Such conditional independence test (per edge) requires inversion of the matrix which is the task of  $O(6^3)$ (single phase) or $O(14^3)$ (three phase) complexity. As mentioned already in the previous section,  an important feature of the test is its independence from the size of the network.

\textbf{Thresholding:} Note that due to numerical errors, empirical estimates of true covariances may not be zero. Thus, we use the following thresholding in the test of the empirical conditional covariance to make decision on  conditional independence of voltages in Algorithm $1$:
\begin{align}
    V^{in}_k\indep V^{in}_l|(V^{in}_i,V^{in}_j) ~~~\text{if}~~~ \left|\left[\Sigma^{in}_{kl,ij}\right]^{-1}_{(1,2)}\right| < \tau_{abs}. \label{cond_abs}
\end{align}
We call this $\text{\textbf{cond}}_\text{\textbf{abs}}$\textbf{-test} with positive threshold $\tau_{abs}$. However voltages at nodes $k,l$ that are far apart may have low correlation and appear uncorrelated given even non-neighbor pair $i,j$. Thus we consider a relative test termed $\text{\textbf{cond}}_\text{\textbf{rel}}$\textbf{-test} with threshold $\tau_{rel}$:
\begin{align}
    V^{in}_k\indep V^{in}_l|(V^{in}_i,V^{in}_j) ~~\text{if}~ \left|\left[\Sigma^{in}_{kl,ij}\right]^{-1}_{(1,2)}\big/\left[\Sigma^{in}_{kl,ij}\right]_{(1,2)}\right|  < \tau_{rel}  \label{cond_ref}
\end{align}

From the graphical model it is clear that removing a single node $i$ or $j$ does not make $k,l$ conditionally independent even if one of them exists in the path from $k$ to $l$. Empirically, however, covariance between $k$ and $l$ after conditioning on one of the two nodes $i,j$ may be significantly reduced despite $i,j$ not being an edge. We thus consider a hybrid test for conditional covariance termed $\text{\textbf{cond}}_\text{\textbf{mod}}$\textbf{-test}, where we also look at the effect of conditioning on both $i,j$ relative to only one of $i$ or $j$:
\begin{align}
    V^{in}_k\indep V^{in}_l|&(V^{in}_i,V^{in}_j) ~\text{if}~  
    \nonumber\\
    &\min\left(\left|\frac{\left[\Sigma^{in}_{kl,ij}\right]^{-1}_{(1,2)}}{\left[\Sigma^{in}_{kl,i}\right]^{-1}_{(1,2)}}\right|, \left|\frac{\left[\Sigma^{in}_{kl,ij}\right]^{-1}_{(1,2)}}{\left[\Sigma^{in}_{kl,j}\right]^{-1}_{(1,2)}}\right|\right)  < \tau_{mod} \label{cond_mod}
\end{align}
An advantage of the relative tests (Eq.~(\ref{cond_ref},\ref{cond_mod})) over the absolute test (Eq.~(\ref{cond_abs})) is through the feature that the thresholds used are less affected by network parameters, nodal injection covariances and other features that can vary within the network.

\begin{figure}[!bt]
\centering
\includegraphics[width=0.29\textwidth,height =.09\textwidth]{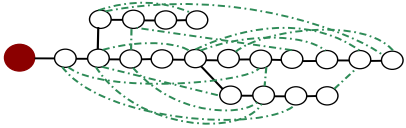}
\vspace{-.10cm}
\caption{Layout  of $20$-bus radial distribution grid. Red circle marks the substation/reference bus. Black lines mark operational edges. The additional permissible edges available to Algorithm $1$ are represented by dotted green lines.}
\label{fig:case19}
\vspace{-2mm}
\end{figure}
\begin{figure}[!ht]
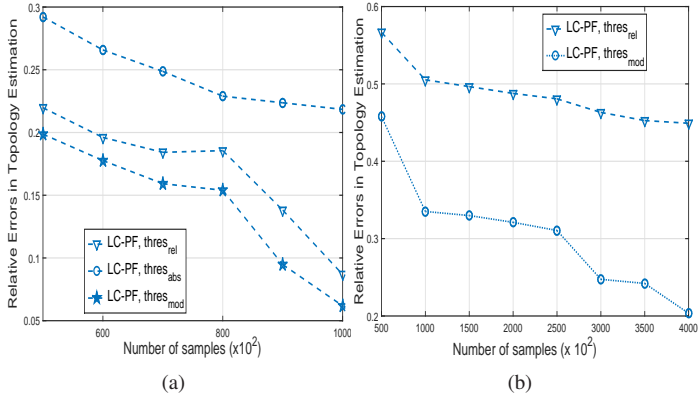

\centering\hfill
\subfigure[]{\includegraphics[width=0.25\textwidth,height = .26\textwidth]{adj_errors_19.eps}\label{fig:adj_errors_19}}
\subfigure[]{\includegraphics[width=0.25\textwidth,height = .26\textwidth]{adj_errors_19_full.eps}\label{fig:adj_errors_19_full}}
\caption{Accuracy of topology learning algorithm with different conditional independence tests and increasing number of LC-PF voltage samples for $20$ bus test case in Fig.~\ref{fig:case19}. (a) $\mathcal{E}_{full}$ has $33$ edges. (b) $\mathcal{E}_{full}$ has $171$ edges (i.e. all non-substation node pairs are considered legitimate).}
\vspace{-3mm}
\end{figure}
\textbf{Comparison of different conditional independence tests:} We discuss the choice of conditional independence test to be used in the Algorithm $1$. We consider a tree distribution network \cite{testcase2,radialsource} with $19$ load nodes and one substation as shown in Fig.~\ref{fig:case19}. We simulate active and reactive load profiles to follow Gaussian random variables uncorrelated across nodes with covariance values around $10\%$ of the load/injections means. We generate nodal complex voltage samples from the Gaussian independent loads/injections through the LC-PF model. The voltage measurements are provided as input together with permissible edge set $\mathcal{E}_{full}$ of $33$ edges ($18$ true edges and $15$ additional edges as shown in Fig.~\ref{fig:case19}). We test the Algorithm $1$ with three threshold-based conditional independence/covariance tests, described above in Eqs~(\ref{cond_abs},~\ref{cond_ref},~\ref{cond_mod}), for sample data set of varying size. The average estimation errors (relative to the number of operational edges) generated by Algorithm $1$ are presented in Fig.~\ref{fig:adj_errors_19}. Note that while increasing the number of samples leads to lesser number of errors for all three tests, the performance of $\text{\textbf{cond}}_\text{\textbf{mod}}$\textbf{-test} (Eq.~(\ref{cond_mod})) is the best at higher sample sizes. This is further demonstrated in Fig.~\ref{fig:adj_errors_19_full} where Algorithm $1$ inputs all the non-substation node pairs (171 in total) as a permissible set of edges. Observe that the performance of $\text{\textbf{cond}}_\text{\textbf{mod}}$\textbf{-test} is better than that of $\text{\textbf{cond}}_\text{\textbf{rel}}$\textbf{-test}. The tolerance values used in the three tests are manually optimized by trial and error. In practice, they can be selected from experiments conducted with historical data. In the simulations section, we use $\text{\textbf{cond}}_\text{\textbf{mod}}$\textbf{-test} for the conditional covariance estimation and edge detection.
\end{document}